\documentclass[multphys,vecphys]{svmult}

\usepackage{makeidx}         
\usepackage{graphicx}        
\usepackage{multicol}        

\makeindex             

\begin{document}

\title*{Hierarchical Formation of Galactic Clusters}
\author{Bruce G. Elmegreen\inst{1}}
\institute{IBM Research Division, T.J. Watson Research Center,
1101 Kitchawan Road, Yorktown Hts., NY 10598, USA
\texttt{bge@watson.ibm.com}}
%
%
\maketitle

Young stellar groupings and clusters have hierarchical patterns
ranging from flocculent spiral arms and star complexes on the
largest scale to OB associations, OB subgroups, small loose
groups, clusters and cluster subclumps on the smallest scales.
There is no obvious transition in morphology at the cluster
boundary, suggesting that clusters are only the inner parts of the
hierarchy where stars have had enough time to mix.  The power-law
cluster mass function follows from this hierarchical structure:
$n(M_{cl})\propto M_{cl}^{-\beta}$ for $\beta\sim2$. This value of
$\beta$ is independently required by the observation that the
summed IMFs from many clusters in a galaxy equals approximately
the IMF of each cluster.

\vspace{.3in}

\noindent From the conference ``Globular Clusters, Guide to
Galaxies,'' 6-10 March 2006, University of Concepcion, Chile, ed.
T. Richtler, et al., ESO/Springer

\section{Introduction}
\label{sec:1}

One way to study the origin of clusters is to observe the general
morphology of star formation, how it fits with the surrounding gas
distribution, and how it responds to various energy sources like
other star formation or galactic-scale processes like spiral
waves. A startling revelation in ISM structure came 20 years ago
following Infrared Astronomical Satellite surveys of dust emission
and CO and HI surveys of the Galactic plane. These surveys
demonstrated that interstellar gas is not a random arrangement of
round clouds with a smooth intercloud medium. It is a scale-free
continuum of structures with a power-law power spectrum,
correlated velocities and densities, and no obvious limits on
either large or small scales except those defined by the galaxy
itself. This view was present as far back as the 1950s, but
dropped from ISM models in the intervening two decades (see review
in \cite{ES04}). Particularly important papers were those by Beech
\cite{B87}, in which fractal structure was found for the first
time in the Lynds dark clouds, Low et al. \cite{L84}, in which the
pervasive and highly structured IR cirrus clouds were discovered,
Crovisier \& Dickey \cite{CD83}, in which the power-law power
spectrum of HI was discovered, and Stutzki et al. \cite{S98}, in
which the power-law power spectrum of CO emission was discovered.
All of these followed the influential paper by Larson \cite{L81}
in which giant molecular cloud (GMC) properties like size,
density, and velocity dispersion were shown to be correlated in a
manner reminiscent of turbulence. Whole galaxies were eventually
found to have correlated properties too \cite{S99,W99,EKS01}.
These papers and others led to a change in thinking about the
structure, energetics, and evolution of interstellar gas. It did
not take long for the theory of star formation to change with it.

Star formation is now seen to follow scale-free patterns like the
gas, suggesting that stars form in turbulent gas wherever the
density is large, making clusters and loose groups with a
power-law distribution of masses. The result is a hierarchy of
clouds and young stellar groupings with HI ``superclouds''
\cite{E79} and star complexes \cite{Ef78,Ef95} on the largest
scale, GMCs and OB associations generally clustered together
inside of them \cite{EE83,EE87,G87}, and molecular cloud cores
with galactic clusters inside. An early review is in Scalo
\cite{S85}. GMCs are not isolated regions of star formation, nor
are they the largest scale of cloud structure. GMCs are not
ballistic objects, nor long-lived objects, although their pieces
may shuffle around and form new clouds after star formation breaks
them apart \cite{E79}. Self-gravitating clouds more massive than
GMCs are observed but they are not molecular. Virialized density
decreases with increasing mass at constant pressure, so the
largest self-gravitating clouds do not self-shield against
dissociative radiation in our Galaxy and are mostly atomic
\cite{EE87}. The largest clouds are more highly molecular in a
high-pressure galaxy like M51 \cite{RK90}. Generally the molecular
fraction in a galaxy follows the pressure \cite{E93,WB02}. Thus
the largest GMC mass in our Galaxy \cite{WM97} is not the largest
self-gravitating cloud mass. Neither is the largest GMC mass
related to the largest cluster mass, which should have a different
dependence on ISM properties like pressure $P$ and core density
$n$: $M_{cl,max}\sim6\times10^3\left(P/3\times10^8\;k_{\rm
B}\right)^{1.5}\left(n/10^5\;{\rm cm}^{-2}\right)^{-2}$
\cite{EE01}.

GMCs are not special cloud structures, they are only the dense
self-shielded parts of the ISM hierarchy \cite{AAT86,EE87}, and
among them, only the most massive tend to be self-gravitating in
large-scale surveys \cite{HCS01}. Molecular cores inside GMCs can
be self-gravitating too. Density peaks in the diffuse ISM should
be viewed as transient, lasting only a few internal turbulent
crossing times before shear and random motions from inside and
outside change their identities.  Even GMCs and their star-forming
cores are probably somewhat transient, although perhaps not for
the same reasons as diffuse clouds. GMC cores appear to begin star
formation very quickly after they form, and the pressure from this
star formation disrupts them \cite{E00,HBB01}. Short cloud
lifetimes appear to be the norm over a wide range of scales, from
individual clusters to whole star complexes, with the actual time
scale for formation increasing with size, in proportion to the
turbulent crossing time \cite{EfE98}.

Hierarchical star formation has been recognized for a long time.
In a series of papers in the 1980's, Feitzinger and collaborators
quantified the hierarchical structure and fractal dimension in
several nearby galaxies, including the LMC \cite{FB84,FG87,FG88}.
Other recent studies of the top two levels in the hierarchy, from
star complexes to OB associations, were made for the LMC
\cite{M98,HZ99,G00}, the SMC \cite{CC05}, M31 \cite{BEM96}, NGC
300 \cite{P01}, M51 \cite{B05}, and M33 \cite{I05}. Large surveys
of galaxies were in Bresolin \cite{B98} and Gusev \cite{G02}.
These studies found collections of OB stars in OB associations,
typically 80 pc in diameter, which are themselves collected into
star complexes several hundred pc in diameter. A review of star
complexes is in Efremov \cite{Ef95}.

The top of the hierarchy consists of $10^7$ M$_\odot$ clouds and
star complexes that are most likely formed by gravitational
instabilities in either the ambient medium, forming flocculent
spirals, or in the dense shocks (dust lanes) of spiral arms when a
stellar spiral wave is strong. Inside density wave spirals, giant
clouds and star complexes are nearly equally spaced with a
separation of about 3 times the arm width \cite{EE83,E06b},
similar to the relative separation of clumps in other filamentary
clouds \cite{SE79}. This is the characteristic length of the
threshold gravitational instability for the tube-like
concentration of gas that is a dust lane \cite{E94}. The
instability also produces feathery clouds that trail off into the
interarm region \cite{B88,KO02}. There might be a characteristic
mass for these largest clouds, comparable to the theoretical Jeans
mass of $\sim c^4/\left(\pi^2G^2\Sigma\right)$ for velocity
dispersion $c$ and gas mass column density $\Sigma$. Such a peaked
mass function is observed in the highly compressed parts of the
interacting galaxy IC 2163 \cite{E06b}, but it is not known
whether such a peaked function is a general feature of giant
spiral arm clouds.

OB associations, loose stellar groupings and star clusters are
evidently fragments of these giant clouds. The associations and
groups usually cluster inside the giant clouds \cite{G87}. On
smaller scales, there are universal power-law mass functions,
which are presumably the result of scale-free processes such as
turbulence and gravitational fragmentation.

The hierarchical structure does not stop at OB subgroups. It
continues down to the sub-parsec scale of individual clusters and
inside the clusters, which are often hierarchical themselves
(e.g., $\rho$ Oph \cite{S05}, NGC 2264 \cite{DS05}, Serpens
\cite{T00}). Kiss \cite{K06} found 3872 T Tauri stars from 2MASS
using colors as a guide. There were 64 possible T associations
among these. These groupings have a star-number distribution that
continues to increase like a power law from 138 stars down to 4
stars. This distribution implies that hierarchical stellar
groupings may continue down to a few stars.

Between the top and bottom of the hierarchy there is a broad range
of correlated scales. As just mentioned, there is hierarchical
structure in stellar groupings, autocorrelation of clusters
\cite{ZFW01}, power-law power spectra of optical light in galaxies
\cite{E03a,E03b,W05}; power-law size distributions of star fields
\cite{EE01,E06c}, and fractional powers in the run of star-counts
versus distance \cite{CC05}. In general, young star fields are
hierarchical, even if the groupings are unbound (e.g. \cite{G00}
for the LMC). Local groupings of low mass x-ray stars (T Tauri
stars) in Gould's Belt also have a hierarchical structure
\cite{G98}.

\section{Cluster Mass Functions}

The power-law mass functions for atomic and molecular clouds and
for star clusters and OB associations are most likely the result
of the hierarchical nature of the ISM. Clouds like these are not
isolated objects but are interconnected by diffuse and molecular
gas in a widespread network. The network has a power-law power
spectrum, as discussed above, and this means there is no
characteristic cloud size or mass. There may be only upper and
lower limits. Because clusters form in scale-free gas clouds with
about constant efficiency (to within a factor of $\sim3$), their
initial stellar masses are also scale free.  The sizes of clusters
are apparently not scale-free as there seems to be a
characteristic cluster radius \cite{Sh06}.

The mass distribution of clusters in spiral galaxy disks is a
power law with a negative slope of around $\beta=2$ on a log-log
plot with linear intervals of cluster mass (or a negative slope of
$1$ on a log-log plot with log intervals of mass). In the Antennae
galaxy, $\beta=1.95\pm0.03$ for young clusters and
$\beta=2.00\pm0.08$ for old clusters \cite{ZF99}. For the LMC,
$\beta=1.85\pm0.05$ \cite{GA06}. For M51, $\beta=2$ \cite{G06}.
For NGC 3310, $\beta=2.04\pm0.23$ and for NGC 6745,
$\beta=1.96\pm0.15$ \cite{G03}. A mass function with a slope of
$\beta=2$ follows from an ISM that is fractal with a
three-dimensional power spectrum slope of $-3.66$
\cite{cancun04,E06c}, the same as for velocity in 3D Kolmogorov
turbulence.

Summed IMFs from clusters can produce a global IMF that is nearly
the same as the individual IMFs if the cluster mass function slope
is the observed value of $\beta=2$.  Galaxy-wide IMFs are in fact
very close to the Salpeter IMF. These IMFs are determined in a
variety of ways, including metallicity, colors, H$\alpha$
equivalent widths, and color-magnitude diagram star counts.
Cluster IMFs have an average slope comparable to the Salpeter
value also \cite{Sc98}. This agreement between galaxy and cluster
IMFs implies that the cluster mass function is close to $\beta=2$.
A slightly steeper cluster mass function makes the summed IMF
significantly steeper \cite{KW03,WK05}, and in clear disagreement
with the IMF observations, reviewed in \cite{E06d}. For example,
$\beta=2.3$ implies the galaxy IMF should have a slope of $-2.9$
\cite{WK05}, which is much steeper than the commonly observed
Salpeter slope of approximately $-2.4$. Evidently, galaxy-wide
IMFs are a sensitive measure of the slope of the mass function of
clusters and of general regions of star formation.

The approximate agreement between summed IMFs and cluster IMFs
implies, in practical terms, that stars of any mass can form in
clusters of any mass. A cluster seems to choose its stars randomly
from a universal IMF. Such random sampling would be illogical if a
cluster were to choose a star more massive than itself, but this
event has negligible occurrence in practice \cite{E06d}.  A
related property is that a power-law IMF implies that the maximum
likely star mass out of N clusters of mass M equals the maximum
likely star mass in one cluster of mass NM \cite{E06d}.

Hierarchical structure with $\beta=2$ is also required by the
observation that cluster IMFs are independent of cluster mass.
More massive clusters have more sub-units for the local sum of
sub-unit IMFs, and this would produce steeper IMFs for more
massive clusters than low mass clusters if $\beta>2$.

\section{Summary}

The hierarchy of star formation extends from ``star complexes" on
$\sim500$ pc scales to the interiors of embedded clusters on
sub-pc scales. This smooth continuation is evident from power
spectra, mass functions, autocorrelation functions, and other
studies. There is apparently no threshold or change at a
``cluster" boundary other than the change from unmixed
hierarchical structure outside the boundary to mixed stellar
orbits inside the boundary. Clusters are best defined dynamically
where the cluster age equals the dynamical time at a density much
higher than the tidal limit. Cluster self-boundedness is a
separate issue. In fact, the efficiency of star formation is
automatically large at high density in a power law ISM
\cite{E02,E05}, so self-boundedness is somewhat inevitable for
very young clusters once the local star formation process ends.

Hierarchical structure produces a cluster mass spectrum with equal
mass in equal logarithmic intervals. This means $\beta\sim2$,
particularly with a Kolmogorov-like power spectrum for structure.
In fact, $\beta\sim2$ is observed directly for many cluster
systems. $\beta\sim2$ is also required by the observation that
galaxy-wide IMFs are equal to individual cluster IMFs. One
important implication of this is that stars of any mass can be
associated with clusters with any number of stars.

The top of the hierarchy (flocculent spiral arms and star
complexes) should have a characteristic cloud or cluster mass
comparable to the ambient Jeans mass, although there are few
observations to confirm this point. A peaked mass function for
star complexes has been observed for only one galaxy so far; the
bound clusters inside these star complexes have a power-law mass
function\cite{E06c}. If halo globular clusters had a peaked mass
function at birth, then it seems possible they were at the top of
a hierarchy of cloud structures also, but in ultra-high pressure
regions to make them compact.



\printindex
\end{document}